\documentclass[prc,twocolumn,showpacs,preprintnumbers,amsmath,amssymb,superscriptaddress,floatfix,nofootinbib]{revtex4}
\usepackage{graphicx}
\usepackage{amsmath}
\usepackage{amsfonts}
\usepackage{amssymb}%

\newcommand{\Slash}[1]{\ooalign{\hfil/\hfil\crcr$#1$}}

\begin{document}

\title{$\Sigma^*_{1/2^-}(1380)$  in the $\Lambda^+_c \to \eta \pi^+ \Lambda$ decay}

\author{Ju-Jun Xie}
\affiliation{Institute of Modern Physics, Chinese Academy of
Sciences, Lanzhou 730000, China}

\author{Li-Sheng Geng} \email{lisheng.geng@buaa.edu.cn}
\affiliation{School of Physics and Nuclear Energy Engineering and
International Research Center for Nuclei and Particles in the
Cosmos and Beijing Key Laboratory of Advanced Nuclear Materials and Physics, Beihang University, Beijing 100191, China}

\date{\today}

\begin{abstract}

A $\Sigma^*$ state with spin-parity $J^P = 1/2^-$  with mass and
width around $1380$ MeV and $120$ MeV, refereed to as the
$\Sigma^*_{1/2^-}(1380)$, has been predicted in several pentaquark
models and inferred from the analysis of CLAS $\gamma p$ data. In
the present work, we discuss how one can employ the $\Lambda^+_c \to
\eta \pi^+ \Lambda$ decay  to test its existence,  as well as to
study the $\Sigma^*(1385)$  state with $J^P = 3/2^+$. Because the
final $\pi^+ \Lambda$ system is in a pure isospin $I = 1$
combination, the $\Lambda^+_c \to \eta \pi^+ \Lambda$ decay can be
an ideal process to study these $\Sigma^*$ resonances. In
particular, we show that the decay angle and energy distributions of the $\pi^+$
are very different for $\Sigma^*(1385)$ and
$\Sigma^*_{1/2^-}(1380)$. The proposed decay mechanism as well as
the existence of the $\Sigma^*_{1/2^-}(1380)$ state  can be checked
by future BESIII and Belle experiments.

\end{abstract}

\pacs{13.75.-n; 14.20.Gk; 13.30.Eg.} \maketitle

\section{Introduction}

Study of the spectrum of $\Sigma^*$ states is one of the most
important issues in hadronic
physics~\cite{Klempt:2009pi,Crede:2013kia}. $\Sigma^*$ states
were mostly produced and studied in $\bar{K}$-induced reactions, and
our knowledge on them  is still rather
limited~\cite{Klempt:2009pi,Crede:2013kia,Olive:2016xmw}. In the
low-lying energy region, only a few $\Sigma^*$ excited states are
well established, such as the $\Sigma^*(1385)$ of spin-parity $J^P =
3/2^+$, $\Sigma^*(1660)$ of $J^P = 1/2^+$, $\Sigma^*(1670)$ of $J^P
= 3/2^-$, $\Sigma^*(1750)$ of $J^P = 1/2^-$, and $\Sigma^*(1775)$ of
$J^P = 5/2^-$. The others are not well established and for some even their existence has not  been confirmed~\cite{Olive:2016xmw}. Thus,
more studies on  $\Sigma^*$ resonances both on theoretical and
experimental sides are necessary.

Based on the penta-quark picture, a new $\Sigma^*$ state with $J^P = 1/2^-$, referred to as the
$\Sigma^*(1380)$,  was predicted with mass around 1380
MeV~\cite{Zhang:2004xt}. Another more general penta-quark
model~\cite{Helminen:2000jb} without introducing explicitly diquark
clusters also predicts this new $\Sigma^*$ state but with mass
around 1405 MeV. The possibility for the existence of such a new
$\Sigma^*(1380)$ state in $J/\psi$ decays was pointed out in
Refs.~\cite{Zou:2006uh,Zou:2007mk}. Later on, the studies of the $K^- p
\to \Lambda \pi^+ \pi^-$ reaction have shown some further evidence
for the existence of the $\Sigma^*(1380)$ state, yielding a mass around
$1380$ MeV and  a width about $120$ MeV~\cite{Wu:2009tu,Wu:2009nw}. Furthermore, in
Refs.~\cite{Gao:2010hy,Chen:2013vxa,Xie:2014zga}, the role played by
the new $\Sigma^*(1380)$ state in the $K \Sigma^*(1385)$
photo-production and $\Lambda p \to p \Lambda \pi^0$ reaction was
studied, and it was shown that, apart from the existing
$\Sigma^*(1385)$ resonance, there are signs of  the $\Sigma^*(1380)$ state. Recently, the existence of an isospin $I=1$
resonance in the vicinity of the $\bar{K}N$ threshold was studied in
Ref.~\cite{Roca:2013cca} based on the analysis of the CLAS data on
the $\gamma p \to K^+ \pi^{\pm} \Sigma^{\mp}$
reactions~\cite{Moriya:2013eb}. Such a sate is also discussed in
Refs.~\cite{Oller:2000fj,Oller:2006jw,Guo:2012vv} within the unitary
chiral perturbation theory. However, the existence of such an $I=1$ state
around the $\bar{K}N$ threshold is less clear since it depends on
the details of the fits performed~\cite{Guo:2012vv}. Clearly, it
is helpful to check the validity of penta-quark models by
studying the contributions of the $\Sigma^*(1380)$ state in different reactions.
Because the mass of this new $\Sigma^*$ state is close to the well
established $\Sigma^*(1385)$ resonance, it will manifest itself in the
production of the $\Sigma^*(1385)$ resonance and as a result an experimental study of
the $\Sigma^*(1385)$ resonance might interfere with that of the $\Sigma^*(1380)$, because their mass overlaps  and they share the same  $\pi \Lambda$ decay mode.

Recently, it has been shown that the non-leptonic weak decays of
charmed baryons are useful processes to study hadronic resonances, some
of which are subjects of intense debate about their
nature~\cite{Klempt:2007cp,Crede:2008vw,Chen:2016qju}. For instance,
the $\Lambda^+_c \to \pi^+ MB$ weak decays were studied in
Ref.~\cite{Miyahara:2015cja}, where $M$ and $B$ stand for
mesons and baryons. It is shown there that these weak
decays might be ideal processes to study the $\Lambda(1405)$ and
$\Lambda(1670)$ resonances, because they are dominated by the isospin $I =
0$ contribution. In Ref.~\cite{Hyodo:2011js}, the $\pi \Sigma$ mass
distribution was studied in the $\Lambda^+_c \to \pi^+ \pi \Sigma$
decay with the aim of extracting the $\pi \Sigma$ scattering
lengths. In a recent work~\cite{Lu:2016ogy} the role of the
exclusive $\Lambda^+_c$ decays into a neutron in testing the flavor
symmetry and final state interactions was investigated. It was shown
that the three body non-leptonic decays are of great interest to
explore  final state interactions in $\Lambda_c^+$ decays. Along
the same line, in Ref.~\cite{Xie:2016evi}, the $\Lambda^+_c \to \pi^+
\eta \Lambda$ decay was revisited taking into account both the $\eta
\Lambda$ and $\pi^+ \eta$ final state interactions. It was found that the $\pi^+ \eta$ and $\eta \Lambda$
invariant mass distributions show clear cusp and peak structures,
which can be associated with the $a_0(980)$ and $\Lambda(1670)$
resonances. These results clearly show that the
$\Lambda_c^+$ decays provide an alternative useful source to obtain
information on the structure of low lying hadronic states.

One should note that the above mentioned
works~\cite{Miyahara:2015cja,Xie:2016evi} considered only the color-favored
external $W$-emission  diagrams, but neglected the color-suppressed $W$-exchange
diagrams~\cite{Lu:2009cm,Cheng:2010vk}. On the other hand, the experimental
measurements of the decay modes of $\Lambda^+_c \to (\pi
\Sigma)^+$, $\eta \Sigma^+$, and $\eta
\Sigma^{*+}$~\cite{Olive:2016xmw,Ammar:1995je} indicate that
the $W$-exchange diagrams, which are subject to color and helicity
suppression, can become relevant in certain $\Lambda^+_c$
decay modes~\cite{Cheng:2015iom}, where the external $W$-emission diagrams do not contribute.
We note that recently the possibility of searching for
$\Xi^0_{bc}$ and $\Xi^+_{cc}$ is explored in the $W$-exchange
processes, $\Xi^0_{bc} \to p K^-$ and $\Xi^+_{cc} \to
\Sigma^{++}_c(2520)K^-$~\cite{Li:2017ndo}.

In this work, we study the role of the $\Sigma^*_{1/2^-}(1380)$ in
 the $\Lambda^+_c \to \eta \Sigma^*_{1/2^-}(1380) \to
\eta \pi^+ \Lambda$ decay, which
can proceed via the external $W$-emission diagram, similar to the $P_c$ states
produced in the $\Lambda^0_b \to K^- P^+_c$
decay~\cite{Aaij:2015tga}. Meanwhile, for comparison, we study the $\Lambda^+_c \to \eta \Sigma^{*+}(1385) \to
\eta \pi^+ \Lambda$ decay, which is dominated by the $W$-exchange
diagram~\cite{Chau:1995gk}.

This article is organized as follows. In Sec.~\ref{Sec:Formalism},
we present the theoretical formalism of the $\Lambda^+_c \to \eta
\pi^+ \Lambda$ decay. Numerical results and discussions are
presented in Sec.~\ref{Sec:Results}, followed by a short summary in Sec. IV.

\section{Formalism} \label{Sec:Formalism}

In this section, we introduce the theoretical formalism and
ingredients to study the $\Lambda^+_c \to \eta \pi^+ \Lambda$ decay.
In the following, we use $\Sigma^*_1$ and $\Sigma^*_2$ to denote the
$\Sigma^*_{1/2^-}(1380)$ state and the $\Sigma^*(1385)$ resonance.

\subsection{Feynman diagrams and decay amplitudes} \label{feylag}

\begin{figure}[htbp]
\begin{center}
\includegraphics[scale=0.8]{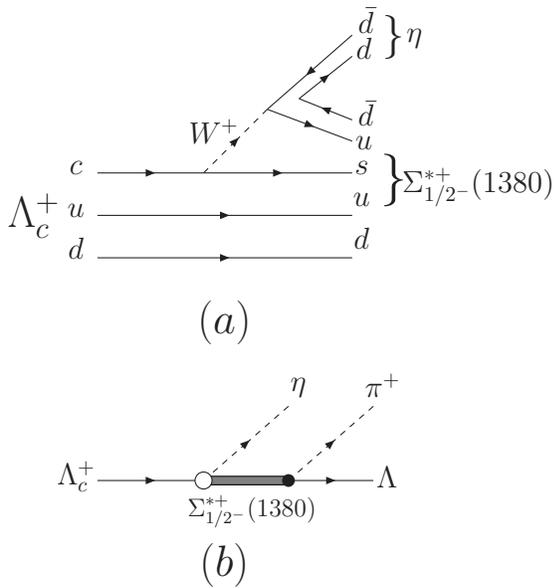}
\caption{Quark level diagram for  $\Lambda^+_c \to \eta
\Sigma^{*+}_{1/2^-}(1380)$  (a) and hadron level diagram for
$\Lambda^+_c \to \eta \Sigma^{*+}_{1/2^-}(1380) \to \eta \pi^+
\Lambda$ decay (b).} \label{Fig:feynd2}
\end{center}
\end{figure}

Because $\Sigma^*_{1/2^-}(1380)$ has a large five-quark
component~\cite{Zhang:2004xt}, it can be produced via the
color-favored  external $W$-emission diagram as shown in
Fig.~\ref{Fig:feynd2} (a). The hadron level diagram for the decay of
$\Lambda^+_c \to \eta \Sigma^{*+}_{1/2^-}(1380) \to \eta \pi^+
\Lambda$ is shown in Fig.~\ref{Fig:feynd2} (b) with
$\Sigma^{*+}(1380)$ decaying into $\pi^+ \Lambda$.

\begin{figure}[t]
\begin{center}
\includegraphics[scale=0.8]{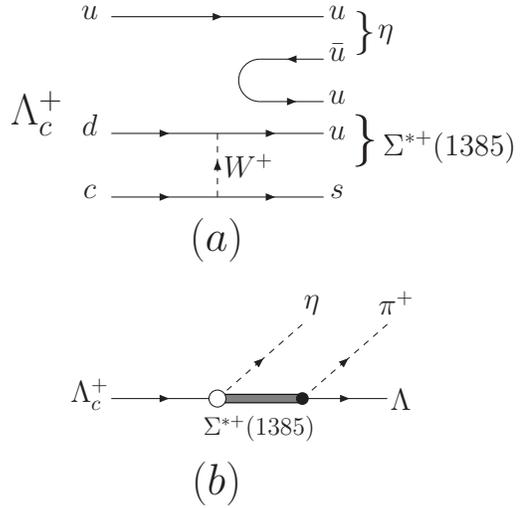}
\caption{Quark  level diagram for $\Lambda^+_c \to \eta
\Sigma^{*+}(1385)$ (a) and hadron level diagram for  $\Lambda^+_c
\to \eta \Sigma^{*+}(1385) \to \eta \pi^+ \Lambda$ decay (b).}
\label{Fig:feynd1}
\end{center}
\end{figure}

The general quark level internal $W$-exchange diagram for the
$\Lambda^+_c \to \eta \Sigma^{*+}(1385)$ is shown in
Fig.~\ref{Fig:feynd1} (a).  In principle, there are also penguin-type quark
diagrams, which, however, can be neglected in charm decays due to
Glashow-Iliopoulos-Maiani cancellation~\cite{Chau:1995gk}. The decay of
$\Lambda^+_c \to \eta \Sigma^{*+}(1385) \to \eta \pi^+ \Lambda$ at
the hadron level is shown in Fig.~\ref{Fig:feynd1} (b).

The general decay amplitudes for $\Lambda^+_c \to \eta
\Sigma^{*+}_{1/2^-}(1380)$ and $\Lambda^+_c \to \eta
\Sigma^{*+}(1385)$ can be decomposed into two different structures
as,
\begin{eqnarray}
{\cal M}(\Lambda^+_c \to \eta \Sigma^{*+}_1) &=& i\bar{u}(q)(A_1 +
B_1 \gamma_5)u(p), \,\\
{\cal M}(\Lambda^+_c \to \eta \Sigma^{*+}_2) &=&
\frac{i}{m_{\eta}}\bar{u}_{\mu}(q)p^{\mu}_1(A_2 + B_2 \gamma_5)u(p),
\,
\end{eqnarray}
where $q$, $p$, and $p_1$ are the momentum of $\Sigma^{*+}_1$ or
$\Sigma^{*+}_2$, $\Lambda_c^+$, and $\eta$ meson, the
$A_1$ and $B_1$ are $s$-wave and $p$-wave amplitudes, 
while $A_2$ and $B_2$ are $p$-wave and $D$-wave amplitudes,
respectively.

To get the whole decay amplitudes of the digrams shown in
Figs.~\ref{Fig:feynd2} (b) and \ref{Fig:feynd1} (b), we use the
interaction Lagrangian densities of
Refs.~\cite{Oh:2007jd,Gao:2012zh,Xie:2013wfa,Xie:2014kja} for
$\Sigma^*_1 \pi \Lambda$ and $\Sigma^*_2 \pi \Lambda$ vertexes,
\begin{eqnarray}
{\mathcal L}_{\pi \Lambda \Sigma^*_1} &=& g_{\pi \Lambda \Sigma^*_1}
\bar{\Sigma}^{*}_1 \vec\tau \cdot \vec\pi \Lambda \, +{\rm h.c.} \,
,  \label{pilambdasigmastar2} \\
{\mathcal L}_{\pi \Lambda \Sigma^*_2} &=& \frac{g_{\pi \Lambda
\Sigma^*_2}}{m_{\pi}} \bar{\Sigma}^{* \mu}_2 (\vec\tau \cdot
\partial_{\mu} \vec\pi) \Lambda \, +{\rm h.c.} \, ,
\label{pilambdasigmastar1}
\end{eqnarray}
where $\Sigma^*_1$ and $\Sigma^{*\mu}_2$ are the fields for
$\Sigma^*(1380)$ and $\Sigma^*(1385)$, respectively.

The coupling constant $g_{\pi \Lambda \Sigma^*_2} = 1.26$ is
determined from the experimental  partial decay width of
$\Sigma^*(1385) \to \pi \Lambda$~\cite{Olive:2016xmw}. For $g_{\pi
\Lambda \Sigma^*_1}$, we fix it to be
$2.12$~\cite{Chen:2013vxa,Xie:2014zga}, assuming that the
$\Sigma^*(1380)$ total decay width, 120 MeV, is solely from the $\pi
\Lambda$ decay.

The invariant decay amplitude of the $\Lambda^+_c \to \eta
\pi^+ \Lambda$ is
\begin{eqnarray}
{\cal M}_{1} & = & i g_{\pi \Lambda \Sigma^*_1} \bar{u}(p_3)
G^{\Sigma^*_1}(q) (A_1 + B_1 \gamma_5) u(p), \label{amplitude1} \\
{\cal M}_{2} & = & \frac{i g_{\pi \Lambda \Sigma^*_2}}{m_{\eta}
m_{\pi}} \bar{u} (p_3) p^{\mu}_2 G^{\Sigma^*_2}_{\mu \nu}(q)
p^{\nu}_1 (A_2 + B_2 \gamma_5) u(p) ,  \label{amplitude2}
\end{eqnarray}
where ${\cal M}_{1}$ and ${\cal M}_{2}$ stand for the contributions
from $\Sigma^*_{1/2^-}(1380)$ and $\Sigma^*(1385)$,
respectively. In the above equations, $p_2$ and $p_3$ represent the
4-momenta of the final $\pi^+$ and $\Lambda$, respectively. The
propagators for$\Sigma^*_{1/2^-}(1380)$ and $\Sigma^*(1385)$  are as
follows,
\begin{eqnarray}
G^{\Sigma^*_1}(q) &=& i \frac{\Slash q + M_{\Sigma^*_1}}{q^2 - M^2_{\Sigma^*_1} + i M_{\Sigma^*_1} \Gamma_{\Sigma^*_1}}, \\
G_{\mu\nu}^{\Sigma^*_2}(q) &=& i \frac{\Slash q +
M_{\Sigma^*_2}}{q^2 - M^2_{\Sigma^*_2} + i
M_{\Sigma^*_2}\Gamma_{\Sigma^*_2}}P_{\mu\nu},
\end{eqnarray}
with
\begin{eqnarray}
P^{\mu \nu} &=& -g^{\mu \nu} + \frac{1}{3}\gamma^{\mu}\gamma^{\nu} +
\frac{2q^{\mu}q^{\nu}}{3M^2_{\Sigma^*_2}} +
\frac{\gamma^{\mu}q^{\nu}-\gamma^{\nu}q^{\mu}}{3M_{\Sigma^*_2}},
\end{eqnarray}
where $M_{\Sigma^*_1}$ ($M_{\Sigma^*_2}$) and $\Gamma_{\Sigma^*_1}$
($\Gamma_{\Sigma^*_2}$) are the mass and total decay width of
$\Sigma^*_{1/2^-}(1380)$ [$\Sigma^*(1385)$] resonance. We take
$M_{\Sigma^*_1}=1380$ MeV and $\Gamma_{\Sigma^*_1}=120$ MeV as in
Refs.~\cite{Wu:2009tu,Wu:2009nw}. For $M_{\Sigma^*_2}$ and
$\Gamma_{\Sigma^*_2}$, we take $M_{\Sigma^*_2} = 1382.8$ MeV and
$\Gamma_{\Sigma^*_2} = 36$ MeV as in the PDG~\cite{Olive:2016xmw}.

\subsection{Invariant mass, decay angle and energy distributions}

The $\pi^+ \Lambda$ invariant mass mass distribution for the $\Lambda^+_c \to
\eta \pi^+ \Lambda$ decay reads~\cite{Olive:2016xmw}
\begin{eqnarray}
\frac{d\Gamma}{dM_{\pi^+ \Lambda}} &=& \frac{m_{\Lambda}}{32 \pi^3
M_{\Lambda^+_c} }  \int \sum |{\cal M}|^2 |\vec{p}_1| |\vec{p}^{~*}|
d{\rm cos}\theta^*,
\end{eqnarray}
where $|\vec{p}^{~*}|$ and $\theta^*$ are the three-momentum and
decay angle of the outing $\pi^+$ (or $\Lambda$) in the
center-of-mass (c.m.) frame of the final $\pi^+ \Lambda$ system,
$|\vec{p}_1|$ is the three-momentum of the final $\eta$ meson in the
rest frame of $\Lambda^+_c$, and $M_{\pi^+ \Lambda}$ is the
invariant mass of the final $\pi^+ \Lambda$ system.

The decay angle and energy distributions of the outgoing particle
can be used to distinguish the intermediate $\Sigma^*$ resonances
with different spin and parity. In the present case, we are
interested in $d\Gamma / d{\rm cos}\theta^*$, which reads
\begin{eqnarray}
\frac{d\Gamma}{d{\rm cos}\theta^*} &=& \frac{m_{\Lambda}}{32 \pi^3
M_{\Lambda^+_c} }  \int \sum |{\cal M}|^2 |\vec{p}_1| |\vec{p}^{~*}|
dM_{\pi^+\Lambda}.
\end{eqnarray} The energy distribution of $\pi^+$ meson reads
\begin{eqnarray}
\frac{d\Gamma}{dE_{\pi^+}} &=& \frac{m_{\Lambda}}{32 \pi^3} \int
\sum |{\cal M}|^2 dE_{\Lambda} ,
\end{eqnarray}
where $E_{\pi^+}$ and $E_{\Lambda}$ are the energies of $\pi^+$ and
$\Lambda$ in the rest frame of $\Lambda^+_c$.

\section{Numerical results and discussion} \label{Sec:Results}

\begin{figure}[htbp]
\begin{center}
\includegraphics[scale=0.45]{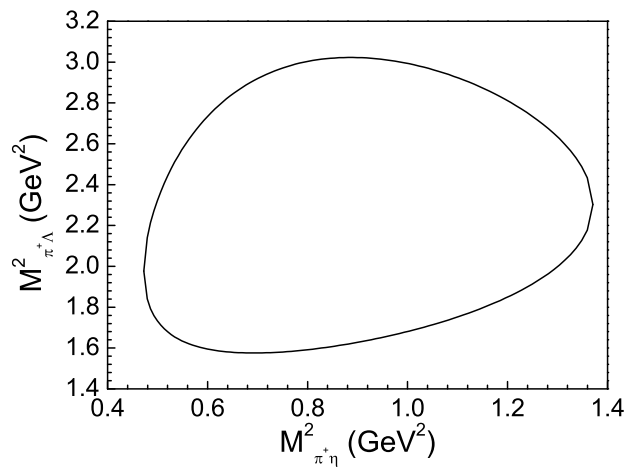}
\caption{Dalitz Plot for $\Lambda^+_c \to \eta \pi^+ \Lambda$ decay,
in the $\pi^+\eta$ and $\pi^+ \Lambda$ invariant masses square.}
\label{Fig:dalitz1}
\end{center}
\end{figure}

In Fig.~\ref{Fig:dalitz1} we show the Dalitz Plot for $M^2_{\eta
\pi^+}$ and $M^2_{\pi^+ \Lambda}$ in the $\Lambda^+_c \to \eta \pi^+
\Lambda$ decay. If we take $M^2_{\pi^+\eta} \sim 1.0$ GeV$^2$, where
the $a_0(980)$ meson gives significant
contributions~\cite{Xie:2016evi}, we see that $M^2_{\pi^+ \Lambda}$
goes from 1.6 GeV$^2$ to 3.0 GeV$^2$, but the range is similar for
other values of $M^2_{\pi^+\eta}$ in a wide range. This means that
the strength of $\pi^+ \Lambda$ invariant mass distribution will
spread in a wide range of $M^2_{\pi^+\eta}$ and we expect that the
contribution from the $a_0(980)$ state will behave roughly like a
background following the phase space. Hence, in this work we do not
consider the contribution from $a_0(980)$ in the calculation of the
$\pi^+ \Lambda$ invariant mass distribution.

\begin{figure}[htbp]
\begin{center}
\includegraphics[scale=0.45]{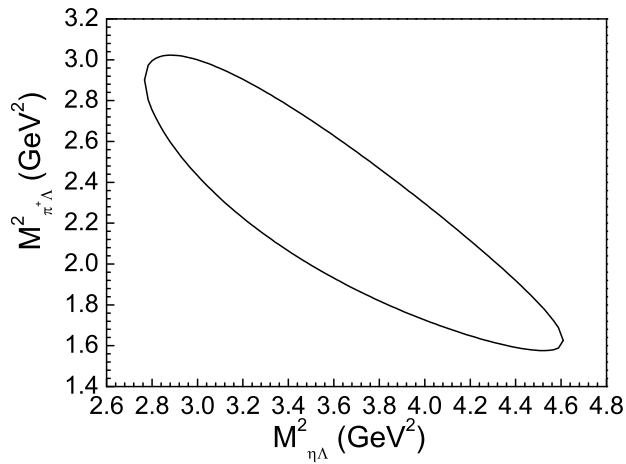}
\caption{Dalitz Plot for $\Lambda^+_c \to \eta \pi^+ \Lambda$ decay,
in the $\eta \Lambda$ and $\pi^+ \Lambda$ invariant masses square.}
\label{Fig:dalitz2}
\end{center}
\end{figure}

In Fig.~\ref{Fig:dalitz2} we show the Dalitz Plot for $M^2_{\eta
\Lambda}$ and $M^2_{\pi^+ \Lambda}$ in the $\Lambda^+_c \to \eta
\pi^+ \Lambda$ decay. If we take $M^2_{\eta \Lambda} \sim 2.8$
GeV$^2$, where the $\Lambda(1670)$ resonance gives significant
contributions~\cite{Xie:2016evi}, we see that $M^2_{\pi^+ \Lambda}$
stays in a very narrow and high energy range from 2.9 GeV$^2$ to 3.0
GeV$^2$, but we are interested in $d\Gamma/dM_{\pi^+ \Lambda}$ in the
range of $M^2_{\pi^+ \Lambda}$ around $1.9$ GeV$^2$. Hence we expect
that the contribution of the $\Lambda(1670)$ resonance will not
affect in any significant way the $\pi^+ \Lambda$ mass distribution
and we neglected its contribution in this work.

In order to evaluate the invariant mass, decay angle, and decay
energy distributions of $d\Gamma/dM_{\pi^+ \Lambda}$, $d\Gamma /
d{\rm cos}\theta^*$ and $d\Gamma/dE_{\pi^+}$ we have to know the
values of $A_1$, $B_1$, $A_2$ and $B_2$. Fortunately, we find that
the shapes of the invariant mass, decay angle, and decay energy
distributions of the $A_1$ ($A_2$) and $B_1$ ($B_2$) terms are
similar and we take $A_1 = B_1$ and $A_2 = B_2$ in this work. They
are also assumed to be constant.~\footnote{In obtaining the decay
amplitude, we have assumed the factorization of the hard process
(the weak decay and hadronization) and the following decays of
$\Sigma^*$ resonances. Such a factorization scheme seems to work
very well (see Ref.~\cite{Oset:2016lyh} for an extensive review). We
note that a combination of the soft-collinear effective theory and
$\chi$PT has been successfully developed to compute the generalized
heavy-to-light form factors~\cite{Meissner:2013hya}, where a similar
factorization scheme is taken but with the hard process calculated
in the QCD perturbation theory.} From the $\Lambda^+_c$ total decay
width $\Gamma_{\Lambda^+_c} = 3.29 \times 10^{-9}$ MeV and the
branch ratio ${\rm Br}[\Lambda^+_c \to \eta \Sigma^{*+}(1385)] =
1.08\%$~\cite{Olive:2016xmw}, we obtain $A_2 = B_2 = 5.51 \times
10^{-7}$, using the following decay width formula
\begin{eqnarray}
&& \Gamma[\Lambda^+_c \to \eta \Sigma^{*+}(1385)] =
\frac{A^2_2|\vec{p}|}{3\pi} \times  \nonumber \\
&& \!\!\! \!\!\! \frac{M^2_{\Lambda^+_c}E^3 - 2 M_{\Lambda^+_c}
M^2_{\Sigma^*_2}E^2 + M^4_{\Sigma^*_2} E -
m^2_{\eta}M^2_{\Sigma^*_2}E}{M_{\Lambda^+_c}m^2_{\eta}M^2_{\Sigma^*_2}},
\end{eqnarray}
with
\begin{eqnarray}
E &=& \frac{M^2_{\Lambda^+_c} + M^2_{\Sigma^*_2} -
m^2_{\eta}}{2M_{\Lambda^+_c}}, \\
|\vec{p}| &=& \sqrt{E^2 - M^2_{\Sigma^*_2}}.
\end{eqnarray}

\begin{figure}[htbp]
\includegraphics[scale=0.45]{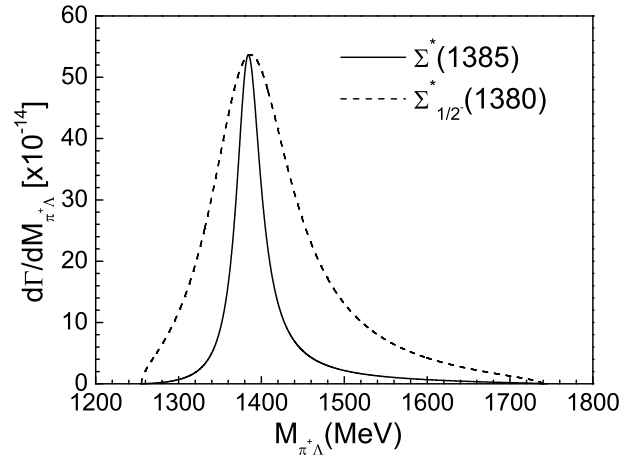}
\caption{Invariant mass distributions $d\Gamma/dM_{\pi^+ \Lambda}$
as a function of $M_{\pi^+ \Lambda}$.} \label{Fig:imd}
\end{figure}

First, we investigate the role of the $\Sigma^*(1385)$ and
$\Sigma^*_{1/2^-}(1380)$ resonances in the invariant mass
distribution of $d\Gamma/dM_{\pi^+ \Lambda}$, which is shown in
Fig.~\ref{Fig:imd}. The solid line stands for the result considering
only the contribution from $\Sigma^*(1385)$
 with $A_2 = B_2 = 5.51 \times 10^{-7}$. While the dashed
curve stands for contributions from only $\Sigma^*_{1/2^-}(1380)$.
For comparison we normalize the two curves to the
peak, which results in $A_1 = B_1 = 13.05 \times
10^{-7}$. From the figure we see that the contribution of
$\Sigma^*_{1/2^-}(1380)$  makes the $\pi^+ \Lambda$ mass
distribution broader because of its relatively large  decay width.

\begin{figure}[htbp]
\includegraphics[scale=0.45]{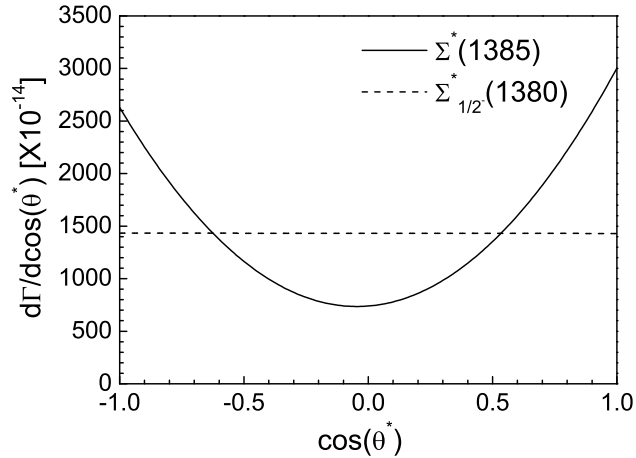}
\caption{Angle distributions $d\Gamma/d{\rm cos}\theta^*$ in the
c.m. frame of $\pi^+ \Lambda$ system as a function of ${\rm
cos}\theta^*$.} \label{Fig:angdis}
\end{figure}

\begin{figure}[htbp]
\includegraphics[scale=0.45]{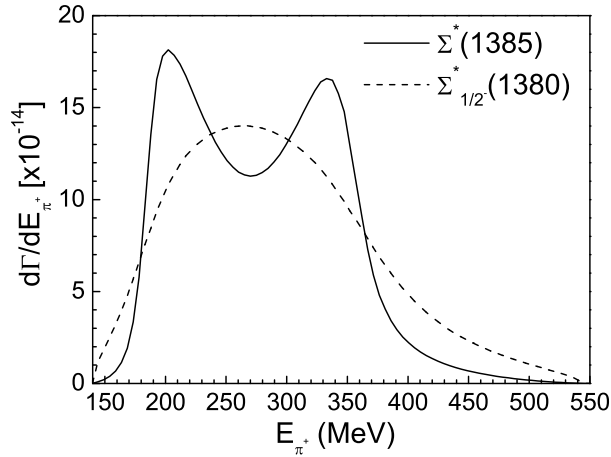}
\caption{Energy distributions $d\Gamma/dE_{\pi^+}$ in the rest frame
of $\Lambda^+_c$ as a function of $E_{\pi^+}$.} \label{Fig:dgde}
\end{figure}

Because the $\Sigma^*(1385)$ resonance has spin-parity $3/2^+$, it
decays into $\pi \Lambda$ in relative $p$-wave, while the
$\Sigma^*_{1/2^-}(1380)$ state with $J^P=1/2^-$ decays into $\pi
\Lambda$ in relative $s$-wave. Hence, we show in
Figs.~\ref{Fig:angdis} and \ref{Fig:dgde}, the decay angle and
energy distributions of the final $\pi^+$, respectively. The solid
and dashed curves stand for the contribution of $\Sigma^*(1385)$ and
$\Sigma^*_{1/2^-}(1380)$, respectively. The two curves are
normalized to the same area in the range examined. One can see that the
shapes of the contributions of $\Sigma^*(1385)$ and
$\Sigma^*_{1/2^-}(1380)$  are very different. From this perspective,
the existence of the $\Sigma^*_{1/2^-}(1380)$ state can be easily
checked by future experimental measurements.

As discussed in the \textit{Introduction}, there is a cusp structure
or a narrow pole near the $\bar{K}N$ threshold in the $I=1$
channel~\cite{Roca:2013cca,Oller:2000fj,Oller:2006jw,Guo:2012vv}.
This structure  may also contribute to the $\Lambda^+_c \to \eta
\pi^+ \Lambda$ decay. However, we expect that its contribution
to the $\pi^+ \Lambda$ invariant mass distribution should be
different from the results shown in Fig.~\ref{Fig:imd}, since the
structure is cusp like  around the $\bar{K}N$ threshold, which could
be easily distinguished from a real resonance.

\section{Summary}

By considering the contributions from the $\Sigma^*(1385)$ and
$\Sigma^*_{1/2^-}(1380)$ resonances, we studied the $\pi^+ \Lambda$
invariant mass, $\pi^+$ decay angle and decay energy distributions
in the $\Lambda^+_c \to \pi^+ \eta \Lambda$ decay to understand
better the $\Sigma^*_{1/2^-}(1380)$ state and also the decay
mechanism. For the production of $\Sigma^*(1385)$, the weak
interaction part is dominated by the internal $W$-exchange diagram,
while for the $\Sigma^*_{1/2^-}(1380)$ production, the weak
interaction part can proceed via the color-favored external
$W$-emission diagram. This is because $\Sigma^*_{1/2^-}(1380)$ has a
dominant
 five-quark component.  The $\Sigma^*(1385)$ and $\Sigma^*_{1/2^-}(1380)$
resonances then decay into a $\pi^+\Lambda$ pair.

As evidenced from the line shape  of the $\pi^+ \Lambda$ invariant
mass distribution, the $\Sigma^*_{1/2^-}(1380)$ state broadens the
invariant mass distribution because of its large total decay width.
Because the $\Sigma^*(1385)$ and $\Sigma^*_{1/2^-}(1380)$ resonances
have different spin and parity, the final $\pi^+$ decay angle and
energy distributions are much different. 

On the experimental side,
the decay mode $\Lambda^+_c \to \pi^+ \eta \Lambda$ has been
observed~\cite{Olive:2016xmw} and the branching ratio
$\mathrm{Br}(\Lambda^+_c \to \pi^+ \eta \Lambda) $ is determined to
be $(2.3 \pm 0.5)\%$, which is one of the dominant decay modes of
the $\Lambda^+_c$ state. Hence, the $\Lambda^+_c \to \pi^+ \eta
\Lambda$ decay can be an ideal process to study the $\Sigma^*(1385)$
and $\Sigma^*_{1/2^-}(1380)$ resonances. Future experimental
measurements of the invariant mass, decay angle and decay energy
distributions studied in the present work will be very helpful in
illuminating the existence of the $\Sigma^*_{1/2^-}(1380)$ state and
improving our knowledge on its properties. For example, a
corresponding experimental measurement could in principle be done by
BESIII~\cite{Ablikim:2015flg} and Belle~\cite{Yang:2015ytm}
Collaborations. Our present study proposed an alternative decay mechanisms for the $\Lambda^+_c \to \pi^+ \eta \Lambda$
decay and constituted a first effort to study the role of the
$\Sigma^*_{1/2^-}(1380)$ state in relevant processes.

\section*{Acknowledgments}

We would like to thank Fu-Sheng Yu and Feng-Kun Guo for useful
discussions. This work is partly supported by the National Natural
Science Foundation of China under Grant Nos. 11475227, 11375024,
11522539, 11505158, and 11475015. It is also supported by the Youth
Innovation Promotion Association CAS (No. 2016367).


\begin{thebibliography}{99}
\bibitem{Klempt:2009pi}
  E.~Klempt and J.~-M.~Richard,
  Rev.\ Mod.\ Phys.\  {\bf 82}, 1095 (2010).
\bibitem{Crede:2013kia}
  V.~Crede and W.~Roberts,
  Rept.\ Prog.\ Phys.\  {\bf 76}, 076301 (2013).
%
\bibitem{Olive:2016xmw}
  C.~Patrignani {\it et al.} [Particle Data Group],
  Chin.\ Phys.\ C {\bf 40}, no. 10, 100001 (2016).
%

\bibitem{Zhang:2004xt}
  A.~Zhang, Y.~R.~Liu, P.~Z.~Huang, W.~Z.~Deng, X.~L.~Chen and S.~L.~Zhu,
  High Energy Phys.\ Nucl.\ Phys.\  {\bf 29}, 250 (2005)
  [hep-ph/0403210].

\bibitem{Helminen:2000jb}
  C.~Helminen and D.~O.~Riska,
  Nucl.\ Phys.\ A {\bf 699}, 624 (2002).

\bibitem{Zou:2006uh}
  B.~S.~Zou,
  Int.\ J.\ Mod.\ Phys.\ A {\bf 21}, 5552 (2006).  

\bibitem{Zou:2007mk}
  B.~S.~Zou,
  Eur.\ Phys.\ J.\ A {\bf 35}, 325 (2008).

\bibitem{Wu:2009tu}
  J.~-J.~Wu, S.~Dulat and B.~S.~Zou,
  Phys.\ Rev.\ D {\bf 80}, 017503 (2009). 
\bibitem{Wu:2009nw}
  J.~-J.~Wu, S.~Dulat and B.~S.~Zou,
  Phys.\ Rev.\ C {\bf 81}, 045210 (2010). 
\bibitem{Gao:2010hy}
  P.~Gao, J.~-J.~Wu and B.~S.~Zou,
  Phys.\ Rev.\ C {\bf 81}, 055203 (2010).
\bibitem{Chen:2013vxa}
  Y.~-H.~Chen and B.~-S.~Zou,
  Phys.\ Rev.\ C {\bf 88}, no. 2, 024304 (2013).

\bibitem{Xie:2014zga}
  J.~J.~Xie, J.~J.~Wu and B.~S.~Zou,
  Phys.\ Rev.\ C {\bf 90}, 055204 (2014).

\bibitem{Roca:2013cca}
  L.~Roca and E.~Oset,
  Phys.\ Rev.\ C {\bf 88}, 055206 (2013).

\bibitem{Moriya:2013eb}
  K.~Moriya {\it et al.} [CLAS Collaboration],
  Phys.\ Rev.\ C {\bf 87}, 035206 (2013).

\bibitem{Oller:2000fj}
  J.~A.~Oller and U.~-G.~Mei\ss ner,
  Phys.\ Lett.\ B {\bf 500}, 263 (2001).

\bibitem{Oller:2006jw}
  J.~A.~Oller,
  Eur.\ Phys.\ J.\ A {\bf 28}, 63 (2006).

\bibitem{Guo:2012vv}
  Z.~H.~Guo and J.~A.~Oller,
  Phys.\ Rev.\ C {\bf 87}, 035202 (2013).

\bibitem{Klempt:2007cp}
  E.~Klempt and A.~Zaitsev,
  Phys.\ Rept.\  {\bf 454}, 1 (2007).

\bibitem{Crede:2008vw}
  V.~Crede and C.~A.~Meyer,
  Prog.\ Part.\ Nucl.\ Phys.\  {\bf 63}, 74 (2009).

\bibitem{Chen:2016qju}
  H.~X.~Chen, W.~Chen, X.~Liu and S.~L.~Zhu,
  Phys.\ Rept.\  {\bf 639}, 1 (2016).


\bibitem{Miyahara:2015cja}
  K.~Miyahara, T.~Hyodo and E.~Oset,
  Phys.\ Rev.\ C {\bf 92}, 055204 (2015).

\bibitem{Hyodo:2011js}
  T.~Hyodo and M.~Oka,
  Phys.\ Rev.\ C {\bf 84}, 035201 (2011).

\bibitem{Lu:2016ogy}
  C.~D.~L\"{u}, W.~Wang and F.~S.~Yu,
  Phys.\ Rev.\ D {\bf 93}, 056008 (2016).


\bibitem{Xie:2016evi}
  J.~J.~Xie and L.~S.~Geng,
  Eur.\ Phys.\ J.\ C {\bf 76}, 496 (2016).

\bibitem{Lu:2009cm}
  C.~D.~Lu, Y.~M.~Wang, H.~Zou, A.~Ali and G.~Kramer,
  Phys.\ Rev.\ D {\bf 80}, 034011 (2009).

\bibitem{Cheng:2010vk}
  H.~Y.~Cheng and C.~W.~Chiang,
  Phys.\ Rev.\ D {\bf 81}, 074031 (2010).

\bibitem{Ammar:1995je}
  R.~Ammar {\it et al.} [CLEO Collaboration],
  Phys.\ Rev.\ Lett.\  {\bf 74}, 3534 (1995).

\bibitem{Cheng:2015iom}
  H.~Y.~Cheng,
  Front.\ Phys.\ (Beijing) {\bf 10}, 101406 (2015).

\bibitem{Li:2017ndo}
  R.~H.~Li, C.~D.~L\"u, W.~Wang, F.~S.~Yu and Z.~T.~Zou,
  Phys.\ Lett.\ B {\bf 767}, 232 (2017).



\bibitem{Aaij:2015tga}
  R.~Aaij {\it et al.} [LHCb Collaboration],
  Phys.\ Rev.\ Lett.\  {\bf 115}, 072001 (2015).


\bibitem{Chau:1995gk}
  L.~L.~Chau, H.~Y.~Cheng and B.~Tseng,
  Phys.\ Rev.\ D {\bf 54}, 2132 (1996).
  
\bibitem{Oh:2007jd}
  Y.~Oh, C.~M.~Ko and K.~Nakayama,
  Phys.\ Rev.\ C {\bf 77}, 045204 (2008). 
\bibitem{Gao:2012zh}
  P.~Gao, J.~Shi and B.~S.~Zou,
  Phys.\ Rev.\ C {\bf 86}, 025201 (2012).
\bibitem{Xie:2013wfa}
  J.~-J.~Xie, B.~-C.~Liu and C.~-S.~An,
  Phys.\ Rev.\ C {\bf 88}, 015203 (2013). 
\bibitem{Xie:2014kja}
  J.~J.~Xie, E.~Wang and B.~S.~Zou,
  Phys.\ Rev.\ C {\bf 90}, 025207 (2014).


\bibitem{Oset:2016lyh}
  E.~Oset {\it et al.},
  Int.\ J.\ Mod.\ Phys.\ E {\bf 25}, 1630001 (2016).

\bibitem{Meissner:2013hya}
  U.~-G.~Mei\ss ner and W.~Wang,
  Phys.\ Lett.\ B {\bf 730}, 336 (2014).

\bibitem{Ablikim:2015flg}
  M.~Ablikim {\it et al.} [BESIII Collaboration],
  Phys.\ Rev.\ Lett.\  {\bf 116}, 052001 (2016).

\bibitem{Yang:2015ytm}
  S.~B.~Yang {\it et al.} [Belle Collaboration],
  Phys.\ Rev.\ Lett.\  {\bf 117}, 011801 (2016).

\end{thebibliography}
\end{document}